\documentclass[aps,prc,superscriptaddress,twoside,twocolumn,nofootinbib,%
showpacs,floatfix,dvips]{revtex4-1}
\usepackage{amsmath,amssymb,graphicx,bm}

\begin{document}

\title{ Photoproduction of
$\gamma N\to K^+ \Sigma^*(1385)$ in the Reggeized framework }


\author{Byung-Geel Yu}
\email[E-mail: ]{bgyu@kau.ac.kr} \affiliation{Research Institute
of Basic Sciences, Korea Aerospace University, Goyang, 412-791,
Korea}
\author{Kook-Jin Kong}
\email[E-mail: ]{kong@kau.ac.kr} \affiliation{Research Institute
of Basic Sciences, Korea Aerospace University, Goyang, 412-791,
Korea}



\begin{abstract}
Photoproduction of  $ K\Sigma^{*}(1385)$ on the  nucleon is
investigated within the Regge framework and the reaction mechanism
is analyzed based on the data existing in the channels $\gamma
p\to K^+\Sigma^{*0}$ and $\gamma n\to K^+\Sigma^{*-}$. The
Reggeization of the $t$-channel meson exchanges
$K(494)+K^*(892)+K_2^*(1430)$ is employed to construct the
photoproduction amplitude. The Rarita-Schwinger formalism is
applied for the spin-3/2$^+$ strangeness-baryon $\Sigma^*$ with  a
special gauge prescription utilized for the convergence of these
reaction processes. Within a set of coupling constants determined
from the symmetry arguement for the $K$ and $K^*$ and from the
duality and vector dominance for the $K_2^*$, the data of the both
processes are reproduced to a good degree. The production
mechanism of these processes are featured by the dominance of the
contact term plus the $K$ exchange with the role of the $K_2^*$
following rather than the $K^*$.
\end{abstract}

\maketitle

\section{Introduction}

Kaon photoproduction off the nucleon target has been a useful tool
to investigate strangeness production with data  on a clean
background from the electromagnetic probe. The experimental
studies of the reactions involving  $\Lambda(1116)$, and
$\Sigma(1190)$ hyperons, or their resonances in the final state
have been extensively conducted up to recent  at the
electron/photon accelerator for hadron
facilities~\cite{mc,glander,bradford,moriya}.

Of recent experimental achievements on these reactions the
measurements of reaction cross sections for the $\gamma p\to
K^+\Sigma^{*0}(1385)$ process from the CLAS \cite{moriya,mattione}
and LEPS \cite{niiyama}, and the $\gamma n\to
K^+\Sigma^{*-}(1385)$ process from the LEPS \cite{hicks}
Collaboration draw our attention.
In these reactions One reason for our interest  is an advantage of
studying baryon resonances whose existences have been predicted by
the quark model, but are still missing, or remain an indefinite
state.
On the other hand,  these reactions have their own issues of how
to deal with the spin-3/2 baryon resonance in describing the
reaction, because the propagation of the spin-3/2 resonance would
give rise to a divergence as the reaction energy increases
\cite{bgyu-pi-delta,bgyu-rho-delta}.

Theoretical investigation of  baryon resonances in the $\gamma
p\to K^+\Sigma^{*0}$ process was carried out  in Ref. \cite{ysoh},
where a set of  $\Delta$ and $N^*$ resonances was considered in
the effective Lagrangian approach. In this pioneering work the
role of the baryon resonances was analyzed  up to spin-5/2 state
in the $s$- and $u$-channel contributions to the reaction process.
Meanwhile, as an extension to the high energy realm a Regge plus
resonance approach was applied for the $\gamma p\to
K^+\Sigma^{*0}$ and $\gamma n\to K^+\Sigma^{*-}$ processes in
Refs. \cite{junhe,wang} with the empirical data up-dated by the
recent experiments. However, in these works, the description of
the reactions was complicated by using a hybrid-type propagation
which mixed the pure Regge-pole and the Feynman propagator in the
$t$-channel, apart from the cutoff functions to suppress the
divergence at high energies, as in Ref. \cite{ysoh}.

In this paper, we investigate photoproduction of $K\Sigma^*$ in
two different isospin channels, $\gamma p \to K^+ \Sigma^{*0}$ and
$\gamma n\to K^+\Sigma^{*-}$, where the Reggeization of the
$t$-channel meson exchange  is exploited for the photoproduction
amplitude at forward angles and high energies.
Our focus here is to describe  these reaction processes up to high
energy without fit-parameters rather than to search for baryon
resonances, because their roles in these reactions are found to be
less important as discussed in Ref. \cite{ysoh}. Avoiding such
complications as mentioned above, we will utilize the model of the
$\gamma N\to \pi^\pm\Delta$ in Ref. \cite{bgyu-pi-delta} to apply
to the present processes with the coupling constant
$f_{KN\Sigma^*}$ considered from the SU(3) symmetry. Since the
$\Sigma^*$ of $3/2^+$ is the lowest mass hyperon in the baryon
decuplet, this will be a valuable test of the flavor SU(3)
symmetry with an expectation that the production mechanism of
$K\Sigma^*$ is essentially identical  to the $\pi\Delta$ case.

For the analysis of the process involving the spin-3/2 baryon
resonance, in particular, it is worth asking how to describe the
process without cutoff functions because they could sometimes hide
the pieces of the reaction mechanism that are missing, or
malfunctioning through the adjustment of the cutoff masses.
From the previous studies on photoproduction of $\pi\Delta$
\cite{bgyu-pi-delta} we have learned two important things as to
the dynamical feature of the spin-3/2 baryon photoproduction: The
minimal gauge prescription is the one requisite for a convergence
of the reaction cross section and the other is the role of the
tensor meson $a_2(1320)$ significant in the high energy region.
Therefore, as a natural extension of the model in Ref.
\cite{bgyu-pi-delta} to strangeness sector, we here consider the
$K(494)+K^*(892)+K_2^*(1430)$~ exchanges in the $t$-channel to
analyze the production mechanism of the $\gamma p\to
K^+\Sigma^{*0}$ and $\gamma n\to K^+\Sigma^{*-}$ processes.

This paper is organized as follows. In Sec.~II, we discuss the
construction of the photoproduction amplitude in association with
the gauge-invariant $K$ exchange in the $t$-channel. This will
include a brief introduction of the minimal gauge, and the new
coupling vertex for the tensor meson interaction $K_2^* N
\Sigma^*$  which has been missed  in previous works. Numerical
results in the total and differential cross sections as well as
the beam polarization asymmetry are presented for  both reactions
in Sec.~III. We give a summary and discussion in Sec.~IV. The
SU(3) coefficients for the octet and decuplet baryons coupling to
octet mesons are given in the Appendix.

\section{formalism}

For a description of the reaction,
\begin{eqnarray}
&&\gamma(k)+N(p)\to K(q)+\Sigma^{*}(p'),
\end{eqnarray}
with the momenta of the initial photon, nucleon and the final $K$
and $\Sigma^*$ denoted by $k$, $p$, $q$, and $p'$, respectively,
we first construct the photoproduction amplitude which is gauge
invariant as to the coupling of photon with particles in the
reaction process.
Then, the Reggeization of the $t$-channel meson-pole follows as
has been done before.

\begin{figure}[t]{}
\centering
\bigskip
\includegraphics[width=0.6\hsize,angle=0]{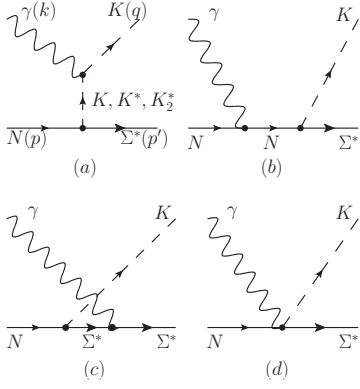}
\caption{Feynman diagrams for $\gamma N \to K^+ \Sigma^{*}$. The
exchange of $K$ in the $t$-channel $(a)$, the proton-pole in the
$s$-channel $(b)$, the $\Sigma^*$-pole in the $u$-channel $(c)$,
and the contact term $(d)$ are the basic ingredients for gauge
invariance of the reaction. The $K^*$ and $K_2^*$ exchanged in the
$t$-channel $(a)$  are themselves gauge-invariant.} \label{fig1}
\end{figure}

\subsection{Photoproduction amplitude}

Viewed from the $t$-channel meson exchange the Born amplitudes in
four different  isospin channels are read as
\begin{eqnarray}
&&{M}_{\gamma p\to K^{+}\Sigma^{*0}}={M}_K+{M}_{K^*}+{M}_{K_2^*},\label{p+}\\
&&{M}_{\gamma n\to
K^{+}\Sigma^{*-}}=\sqrt{2}\left({M}_K+{M}_{K^*}+{M}_{K_2^*}\right),\label{n+}\\
&&{M}_{\gamma p\to K^{0}\Sigma^{*+}}=-\sqrt{2}\left({ M}_{K^*}+{M}_{K_2^*}\right),\label{p0}\\
&&{M}_{\gamma n\to
K^{0}\Sigma^{*0}}=-\left({M}_{K^*}+{M}_{K_2^*}\right),\label{n0}
\end{eqnarray}
where the $\sqrt{2}$ factors and signs result from our convention
of the meson-baryon-decuplet coupling of the $10-8-8$ type
presented in the Appendix.
Hereafter, we call the reaction process in Eq. (\ref{p+}), the
$\gamma p$ process, and the process in Eq. (\ref{n+}), the $\gamma
n$ process, respectively.

In experimental sides, the cross sections for total and
differential were measured recently for the charged state Eq.
(\ref{p+}) at the CLAS \cite{moriya} and LEPS \cite{niiyama}
Collaborations, and the differential cross section and the beam
asymmetry were measured for the process in Eq. (\ref{n+}) at the
LEPS Collaboration \cite{hicks}. There exist data from the CBCG
\cite{crouch} and ABBHHM Collaboration \cite{erbe-nc,erbe-pr} in
the pre-1970's where the total cross section for the charged
process in Eq. (\ref{p+}) as well as the total and differential
cross sections for the process in Eq. (\ref{n+}) reported by the
ABHHM Collaboration \cite{benz}.
Therefore, these data will be of use to constrain the physical
quantities such as the coupling constants in the reaction once the
trajectories of the Regge-poles for $K$, $K^*$, and $K_2^*$ are
chosen.

\subsection{$K(494)$ exchange}

For nucleon, kaon, and $\Sigma^*$ charges, the current
conservation following  the  charge conservation,
$e_N-e_{K}-e_{\Sigma^*}=0$, requires that the $\gamma p$ process
includes the proton-pole in the $s$-channel and the contact term
for gauge-invariance of the $t$-channel $K$ exchange. Similarly
the $\gamma n$ process includes the $u$-channel $\Sigma^*$-pole
and the contact term in addition to the $K$ exchange,
respectively. These are depicted in Fig. \ref{fig1}. Thus, the
gauge-invariant $K$ exchange in the $t$-channel for these
reactions are given by
\begin{eqnarray}\label{pk+}
&&iM_K^{\gamma p}=\overline{u}_{\nu}(p')
i\left[{M}^{\nu\mu}_{t(K)}+M^{\nu\mu}_{s(N)}+
{M}^{\nu\mu}_c\right]
\epsilon_\mu(k)u(p),\\
&&iM_K^{\gamma
n}=\overline{u}_{\nu}(p')i\left[{M}^{\nu\mu}_{t(K)}+
M^{\nu\mu}_{u(\Sigma^*)}+ {M}^{\nu\mu}_c\right]
\epsilon_\mu(k)u(p),\label{nk+}\ \ \
\end{eqnarray}
where
\begin{eqnarray}
&&i{M}^{\nu\mu}_{s(N)}= \Gamma_{K N\Sigma^*}^{\nu}(q)
\frac{\rlap{/}p+\rlap{/}k+M_N}{s-M^{2}_{N}}\Gamma^\mu_{\gamma NN}(k),\label{nucleon}\\
&& i{M}^{\nu\mu}_{t(K)} =\Gamma^{\mu}_{\gamma KK}(q,Q)
\frac{1}{t-m^{2}_{K}} \Gamma_{K N\Sigma^*}^{\nu}(Q),
\label{kaon}\\
&&i{M}^{\nu\mu}_{u(\Sigma^*)}=
\Gamma_{\gamma\Sigma^*\Sigma^*}^{\nu\mu\sigma}(k)\frac{\rlap{/}p'-\rlap{/}k+M_{\Sigma^*}}
{u-M^2_{\Sigma^*}}\Pi^{\Sigma^*}_{\sigma\beta}(p'-k)\nonumber\\&&\hspace{1.5cm}\times
\Gamma_{K N\Sigma^*}^{\beta}(q), \label{sigma33}\hspace{0.5cm}
\end{eqnarray}
with $Q^\mu=(q-k)^\mu$,  the $t$-channel momentum transfer, and
the spin-3/2 projection which is given by
\begin{eqnarray}\label{sigma-propagator}
\Pi_{\Sigma^*}^{\mu\nu}(p)=-g^{\mu\nu}+\frac{\gamma^{\mu}\gamma^{\nu}}{3}
+\frac{\gamma^{\mu}p^{\nu}-\gamma^{\nu}p^{\mu}}{3M_{\Sigma^*}}
+\frac{2p^{\mu}p^{\nu}}{3M^{2}_{\Sigma^*}}\,.\ \
\end{eqnarray}

Here $u_\nu(p')$, $u(p)$, and $\epsilon_\mu(k)$ are the spin-3/2
Rarita-Schwinger field for the $\Sigma^*(1385)$, Dirac spinor for
nucleon, and the spin polarization of photon, respectively.

The charge-coupling vertices $\gamma NN$, $\gamma\Sigma^*\Sigma^*$
and $\gamma KK$ \cite{bgyu-pi-delta} are given as follows,
\begin{eqnarray}
&&\epsilon_\mu\Gamma^\mu_{\gamma
NN}=e_N\rlap{/}\epsilon,\label{gammann}\\
&&\epsilon_\mu\Gamma_{\gamma\Sigma^*\Sigma^*}^{\nu\mu\sigma}
    =-e_{\Sigma^*}\left(g^{\nu\sigma}\rlap{/}\epsilon-\epsilon^\nu\gamma^\sigma
    -\gamma^\nu \epsilon^\sigma +\gamma^\nu\rlap{/}\epsilon
    \gamma^\sigma\right),\label{gamma33}\\
&&\epsilon_\mu\Gamma^{\mu}_{\gamma KK}(q,Q)=e_K (q+Q)^\mu
\epsilon_\mu\,,   \label{gkk}
\end{eqnarray}
where $e_N$, $e_{\Sigma^*}$, and $e_K$  are the nucleon,
$\Sigma^*$ and kaon charges, respectively.

For the strong coupling vertex $K N\Sigma^*$ we use
\begin{eqnarray}
\Gamma^\nu_{K N\Sigma^*}(q)={f_{K N\Sigma^*}\over m_K}q^\nu,
\end{eqnarray}
and neglect the off-shell effect of the spin-3/2 Rarita-Schwinger
field for simplicity. Then, the contact term is given by
\begin{eqnarray}\label{gau3}
i{M}_c^{\nu\mu}=-e_K\frac{f_{K N\Sigma^*}}{m_K}\ g^{\nu\mu}\,.
\end{eqnarray}

Note that the charge-coupling terms in Eqs. (\ref{gammann}),
(\ref{gamma33}), and (\ref{gkk}) satisfy the Ward identities in
their respective vertices \cite{bgyu-pi-delta}, and the full
expressions for the spin-3/2 baryon electromagnetic form factors
will be found in Ref. \cite{bgyu-rho-delta}.

Since the mass of $\Sigma^*$ lies below $\bar{K}N$ threshold the
empirical decay channel $\Sigma\to \bar{K}N$ is not available for
the estimate of the $KN\Sigma^*$ coupling constant, and we follow
the SU(3) symmetry which predicts,
\begin{equation}\label{su3-rel}
\frac{f_{\pi^- p \Delta^{++}}}{m_\pi} = -\sqrt6
\frac{f_{K^+p\Sigma^{*0}}}{m_K}\,,
\end{equation}
and determine the coupling constant $f_{K^+p\Sigma^{*0}}$ from the
empirically known coupling constant $f_{\pi^- p \Delta^{++}}$.
(See  the  Clebsch-Gordan coefficients with phase for the SU(3)
baryon decuplet in the Appendix.) Hereafter, we will write
$f_{K^+p\Sigma^{*0}}$  as $f_{KN\Sigma}$ for brevity.
In our previous work \cite{bgyu-pi-delta} we considered the
coupling constant in the range from $f_{\pi^- p \Delta^{++}}= 1.7$
to $2$. From these we estimate $f_{KN\Sigma^{*}} = -2.46$ and
$-2.83$, respectively.
In other model calculations, however,  the determination of
$f_{KN\Sigma^{*}}$ is found to be rather scattered, e. g.,
$f_{KN\Sigma^{*}}=-3.22$ for the $\gamma p$ process in the
effective Lagrangian approach by applying $f_{\pi^-
p\Delta^{++}}=2.23$ to the symmetry relation above \cite{ysoh}.
The coupling constant $f_{KN\Sigma^*}=-4.74$ for the $\gamma p$
process \cite{junhe}, and $-1.22$  for the $\gamma n$ process
\cite{wang} were obtained from the $\chi^2$-fit of data in the
Regge plus resonance approach. In this work we take the coupling
constant $f_{KN\Sigma^{*}}=-2.2$ within the range discussed above
for a better agreement with experiment.

\subsubsection*{Minimal gauge}

It is well known that the propagation of spin-3/2 $\Sigma^*$
baryon in Eq. (\ref{sigma33}) causes divergence of the reaction at
high energy. However, if we expect that only the peripheral $K$
exchange in the $t$-channel should dominate at high energies and
small angles, then the particle exchanges in the reaction should
contribute only to the Coulomb component of the photoproduction
currents in Eqs. (\ref{pk+}) and (\ref{nk+}). This we call the
minimal gauge prescription for the $K$ exchange advocated in Refs.
\cite{stichel,clark,bgyu-pi-delta}, and this is physically
sensible  because the higher multipoles of the $\Sigma^*$ as a
resonance are defined uniquely in the static limit and such a
uniqueness can no longer be valid at high energy.

In the Reggeized model we recall that the $u$-channel
$\Sigma^*$-pole in Eq. (\ref{nk+}) as well as the $s$-channel
proton-pole term in Eq. (\ref{pk+}) is introduced merely to
preserve gauge invariance for the $t$-channel $K$-pole exchange,
respectively.  By the above speculation at high energies we
consider only the Coulomb components of the $s$-, and $u$-channel
amplitudes that are indispensable to restore gauge invariance of
the $K$ exchange.
Technically speaking, these correspond to the non-gauge invariant
terms in the $s$- and $u$-channels after we remove the transverse
component of the production current by redundancy with respect to
gauge invariance.

In the $u$-channel amplitude in Eq. (\ref{sigma33}), for instance,
the full expression is now written as \cite{bgyu-pi-delta},
\begin{eqnarray}\label{uch1}
&&i{M}_{u(\Sigma^*)} =e_{\Sigma^*} \frac{f_{K N\Sigma^*}}{m_{K}}
\bar{u}^{\nu}(p')\biggl[{ 2\epsilon \cdot p'\over
u-M^2_{\Sigma^*}}g_{\nu\alpha}\nonumber\\&&\hspace{2cm}+G_{\nu\alpha}(p',k)\biggr]q^\alpha
u(p),
\end{eqnarray}
where $G_{\nu\alpha}(p',k)$ is the part of the amplitude which
collects all the terms that are gauge-invariant themselves. Thus,
in this minimal gauge the production amplitude simply consists of
the non-invariant terms in three channels, i.e.,
\begin{eqnarray}\label{min}
&&i{M}_{K}=\frac{f_{K N\Sigma^*}}{m_{K}}
\bar{u}_{\nu}(p')\biggl[q^\nu\frac{2p\cdot\epsilon}{s-M^{2}_{N}}e_N
\nonumber\\&&\hspace{0cm}
+e_{\Sigma^*} \frac{2p'\cdot\epsilon}
{u-M^{2}_{\Sigma^*}}q^{\nu}
%
+ e_{K}\frac{2q\cdot \epsilon }{t-m^{2}_{K}}(q-k)^\nu\biggr]
u+iM_c\,.\hspace{0.3cm}
\end{eqnarray}

With the $K$ exchange given in Eq. (\ref{min}), we now make it
Reggeized by the following procedure,
\begin{eqnarray}\label{regge}
&&{\cal M}_K=M_{K}\times(t-m_K^2){\cal R}^{K}(s,t),
\end{eqnarray}
where
\begin{eqnarray}\label{regge3}
{\cal R}^\varphi
=\frac{\pi\alpha'_\varphi}{\Gamma(\alpha_\varphi(t)+1-J)}
\frac{\mbox{phase}}{\sin\pi\alpha_\varphi(t)}
\left(\frac{s}{s_0}\right)^{\alpha_\varphi(t)-J},\hspace{0.3cm}
\end{eqnarray}
is the Regge pole written collectively for the meson
$\varphi(=K,\,K^*,\,K^*_2)$  of spin-$J$ with the canonical phase
${1\over2}((-1)^J+e^{-i\pi\alpha_J(t)})$ taken for the
exchange-nondegenerate meson in general.

For the trajectory of $K$ we use
\begin{eqnarray}\label{regge2}
&&\alpha_K(t)=0.7\,(t-m_K^2)\,.
\end{eqnarray}

The phases of the $K$ exchange is taken from the reaction $\gamma
p\to K^+\Lambda$ \cite{bgyu-kaon} as a natural extension. As for
the $\gamma n\to K^+\Sigma^-$ process, however, we favor to choose
the phase of the $K$ exchange for a better description of the
reaction processes, as will be discussed later.

\subsection{$K^*(892)$ exchange}

The $K^*$ exchange in the $t$-channel is one of the ingredients to
consider for the analysis of the production mechanism.

The production amplitude is give by \cite{bgyu-rho-delta},
\begin{eqnarray}\label{amp2}
&&i{\cal M}_{K^*}=-i{g_{\gamma KK^*}\over m_0}
\,\epsilon^{\mu\rho\lambda\alpha}
\epsilon_{\mu}k_{\rho}q_{\lambda}
\bar{u}_{\nu}(p')\nonumber\\&&\hspace{1cm}\times\left(-g_{\alpha\beta}+Q_\alpha
Q_\beta/m^2_{K^*}\right)\Gamma^{\beta\nu}_{K^* N\Sigma^*}(Q,p',p)
u(p)\nonumber\\&&\hspace{1cm}\times{\cal R}^{K^*}(s,t).
\end{eqnarray}
For the $K^*N\Sigma^*$ coupling we consider only the following
form,
\begin{eqnarray}\label{kstar}
&&\Gamma^{\beta\nu}_{K^* N\Sigma^*}(q,p',p)={f_{K^*N\Sigma^*}\over
m_{K^*}}\left( q^\beta\gamma^\nu - \rlap{/}q g^{\beta\nu}
\right)\gamma_5,
\end{eqnarray}
and disregard the other nonleading terms  simply because the
leading contribution of the $K^*$ exchange in Eq. (\ref{kstar})
itself is not significant.
In our previous work on the $\gamma p\to\pi^\pm\Delta$ process we
used $f_{\rho N\Delta}=5.5$ for the Model I, and 8.57 for the
Model II \cite{bgyu-rho-delta}. These values lead to
$f_{K^*N\Sigma^*}=-2.58$ and $-4.03$, respectively, according to
the SU(3) relation
\begin{equation}
\frac{f_{\rho N \Delta}}{m_\rho} = -\sqrt6 \frac{f_{K^* N
\Sigma^*}}{m_{K^*}}\,.
\end{equation}
With these values  we try  to find which one yields the better
result in the numerical analysis.
From the decay width $\Gamma_{K^*\to K\gamma}=50$ keV for the
charged state, we estimate $g_{\gamma K^*K}=\pm0.254$ and the take
the negative sign for an agreement with data.

The trajectory for $K^*$ is taken to be
\begin{eqnarray}\label{regge2}
&&\alpha_{K^*}(t)=0.83\,t+0.25 \,,
\end{eqnarray}
which is consistent with the previous works \cite{glv,bgyu-kaon}.
The complex phase  for the $\gamma p$~ and the constant phase for
$\gamma n$ processes are considered for the exchange-degenerate
(EXD) pair $K^*$-$K_2^*$.

\subsection{$K^*_2(1430)$ exchange}

It is found that the tensor-meson $a_2(1320)$ of spin-2 exchange
plays the role at high energy from the previous studies of the
reactions $\gamma N\to\pi^\pm N$ \cite{bgyu-pion} and $\gamma
N\to\pi^\pm\Delta$ \cite{bgyu-pi-delta}. Furthermore the role of
the $K_2^*$ in the strangeness sector is also noticeable in the
$\gamma p\to K^+\Lambda$ \cite{bgyu-kaon}. Therefore, it is quite
reasonable to consider the tensor meson $K_2^*$ exchange in these
reaction processes. As an application of the $a_2N\Delta$ coupling
in Ref. \cite{bgyu-pi-delta} to the strangeness sector, we write
the Lagrangian for the $K_2^*N\Sigma^*$ as,
\begin{eqnarray}
{\cal L}_{K^*_2N\Sigma^*}=i{f_{K_2^*N\Sigma^*}\over
m_{K_2^*}}\overline{\Sigma^*}^\lambda(g_{\lambda\mu}\partial_\nu
+g_{\lambda\nu}\partial_\mu) \gamma_5 N
{K_2^*}^{\mu\nu}.\hspace{0.4cm}
\end{eqnarray}
Here ${K_2^*}^{\mu\nu}$ is the tensor field of spin-2 with the
coupling constant assumed to be
\begin{eqnarray}\label{tensorcc}
{f_{K^*_2 N\Sigma^*}\over m_{K^*_2}}\approx-3{f_{K^*
N\Sigma^*}\over m_{K^*}}
\end{eqnarray}
by a simple extension  to the strangeness sector from the $\rho$
and $a_2$ meson case which is based on the duality and vector
dominance \cite{goldstein,thews}.
In the $\pi\Delta$ photoproduction the tensor meson-$\Delta$
baryon coupling constant determined by such a relation above
yielded a reasonable result in the high energy region, as
illustrated in Ref. \cite{bgyu-pi-delta}.

The Lagrangian for the  $\gamma KK^*_2$ coupling was investigated
in Ref. \cite{giacosa} and   given by
\begin{eqnarray}
{\cal L}_{\gamma KK^*_2}=-i\frac{g_{\gamma
KK^*_2}}{m^2_0}\tilde{F}_{\alpha\beta}(\partial^\alpha
{K_2^*}^{\beta\rho}-\partial^\beta
{K_2^*}^{\alpha\rho})
\partial_\rho K \,,\
\end{eqnarray}
where
$\tilde{F}_{\alpha\beta}={1\over2}\epsilon_{\mu\nu\alpha\beta}F^{\mu\nu}$
is the pseudotensor field of photon. The decay of the tensor meson
$K^*_2$ to $K\gamma$ is reported to be $\Gamma_{K^*_2\to
K\gamma}=(0.24\pm0.05)$ MeV in the Particle Data Group (PDG) and
we estimate $g_{\gamma KK^*_2}=0.276$ \cite{bgyu-kaon} with the
sign determined to agree with existing data.

The Reggeized amplitude for the $K_2^*$ exchange is thus written
as
\begin{eqnarray}
\label{amp4}
&& i{\cal M}_{K_2^*}= -i\frac{2g_{\gamma
KK^*_2}}{m_0^2}{f_{K_2^* N\Sigma^*}\over m_{K_2^*}}
\,\epsilon^{\alpha\beta\mu\lambda}\epsilon_\mu k_\lambda Q_\alpha
q_\rho \nonumber\\
&&\hspace{1cm}\times\Pi_{K_2^*}^{\beta\rho;\sigma\xi}(Q)
\bar{u}^{\nu}(p')(g_{\nu\sigma}P_\xi+g_{\nu\xi}P_\sigma)\gamma_5
u(p)\nonumber\\
&&\hspace{1cm}\times {\cal R}^{K_2^*}(s,t) \,,
\end{eqnarray}
where $P=(p+p')/2$ and the spin-2 projection is given by
\begin{eqnarray}
\Pi^{\beta\rho;\sigma\xi}_{K_2^*}(Q)={1\over2}(\eta^{\beta\sigma}\eta^{\rho\xi}
+\eta^{\beta\xi}\eta^{\rho\sigma})-{1\over3}\eta^{\beta\rho}\eta^{\sigma\xi}
\end{eqnarray}
with $\eta^{\beta\rho}=-g^{\beta\rho}+Q^\beta Q^\rho/m^2_{a_2}$.

For the $K_2^*$ Regge-pole exchange we take the EXD phase
$e^{-i\pi\alpha_{K_2^*}}$ for the $\gamma p$~ and the constant
phase for the $\gamma n$~ processes, respectively, as discussed
above, and choose the trajectory
\begin{eqnarray}\label{regge2}
&&\alpha_{K^*_2}(t)=0.83\,(t-m^2_{K^*_2})+2 \,,
\end{eqnarray}
to be consistent with Ref. \cite{bgyu-kaon}.

\begin{table}[]
\caption{Physical constants from the SU(3) symmetry, Regge
trajectories and phases for $^{(a)}\gamma p\to K^+\Sigma^{*0}$ and
$^{(b)}\gamma n\to K^+\Sigma^{*-}$. The radiative decay constants
are $g_{\gamma KK^*}=-0.254$ and $g_{\gamma KK^*_2}=0.276$.}
    \begin{tabular}{c|c|c|c}\hline
        Meson &  $^{(a)}$Phase  &$^{(b)}$Phase &Cpl. const.  \\
        \hline\hline
        $K$ &  $e^{-i\pi \alpha}$& $(1+e^{-i\pi \alpha})/2$& $f_{KN\Sigma^*}=-2.2$  \\%
        $K^*$ & $e^{-i\pi \alpha}$& $1$ & $f_{K^*N\Sigma^*}=-4.03$  \\%
        $K_2^*$&$e^{-i\pi \alpha}$& $1$ & ${\rm Eq}.~ (\ref{tensorcc})$   \\%
        \hline
        \hline
    \end{tabular}\label{tb1}
\end{table}

In model calculations where the Regge-poles are employed to
estimate physical observables the results in shape and magnitude
are, in general, very sensitive to a change of the phase as well
as the trajectory. Therefore, it is of importance to choose the
phase of $K$ exchange which dominates over other meson exchanges.
We take the complex phase for the $K$ exchange in the $\gamma p$
process, as before. In the case of $\gamma n$ process, however,
the choice of the constant phase leads to an overestimation of the
total cross section in the resonance peak, while fixing the
coupling constant
$f_{K^+n\Sigma^{*-}}=\sqrt{2}f_{K^+p\Sigma^{*0}}$. Without
altering the coupling constant, thus, we take the canonical phase
which is more adaptive to describe the  reaction processes.

In Table \ref{tb1} we list the coupling constants and phases used
for the calculation of the $\gamma p\to K^+\Sigma^{*0}$ and
$\gamma n\to K^+\Sigma^{*-}$ reactions.

\section{numerical Results}

In this section we present numerical consequences in the cross
sections for the total, differential and beam polarization for the
reactions $\gamma p\to K^+\Sigma^{*0}$ and $\gamma n\to
K^+\Sigma^{*-}$.

\subsection{$\gamma p\to K^+\Sigma^{*0}$}

\begin{figure}[]
\centering
\includegraphics[width=0.9\hsize,angle=0]{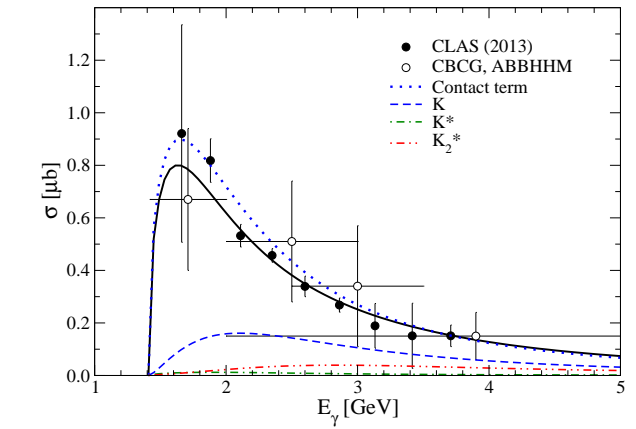}
\caption{Total cross section for $\gamma p \to K^+
\Sigma^{*0}(1385)$. Contributions of the contact term and meson
exchanges are shown in the dotted, dashed, dash-dotted and
dash-dot-dotted curves, respectively. The dominance of the contact
term is shown. The CLAS data are taken from Ref. \cite{moriya}.
The data of the CBCG and ABBHHM are from
Refs.~\cite{crouch,erbe-nc,erbe-pr}. } \label{fig2}
\end{figure}

Given the production amplitudes in Eq. (\ref{p+}) with the
coupling constants in Table \ref{tb1} determined from the symmetry
consideration, we calculate total and differential cross sections
for $\gamma p \to K^+ \Sigma^{*0}$ and present the result to
compare with existing data. There is a discrepancy between the
recent CLAS data and old ones measured in the pre-1970's by the
CBCG \cite{crouch} and ABBHHM Collaboration
\cite{erbe-nc,erbe-pr}. The solid curve in Fig. \ref{fig2}
corresponds to the full calculation of the cross section  with the
coupling constants chosen to agree with the CLAS data, and the
respective contributions of meson exchanges are displayed. As
shown in the figure the production mechanism is solely understood
as the dominating role of the contact term in Eq. (\ref{gau3})
plus the pseudoscalar $K$ exchange in Eq. (\ref{kaon}), while the
tensor meson $K_2^*$ exchange in Eq. (\ref{amp4}) gives a
contribution gradually growing as the energy increases. The
contribution of the vector meson $K^*$ exchange in Eq.
(\ref{amp2}) is small and less significant than that of the
tensor-meson $K_2^*$.
That the $K^*$ contribution is small and thus insignificant is
consistent with the observation in other model calculations of the
process \cite{ysoh}, and confirms the validity of the leading
$K^*N\Sigma^*$ interaction considered only for the $K^*$ exchange.

\begin{figure}[]
\centering \vspace{0.5cm}
\includegraphics[width=0.9\hsize,angle=0]{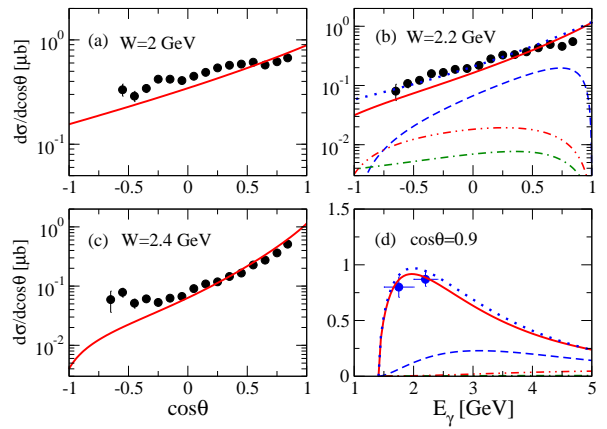}
\caption{Differential cross sections for $\gamma p \to K^+
\Sigma^{*0}(1385)$. The angle dependence of the cross sections are
shown in the first three panels (a), (b), (c) with the data taken
from the CLAS Collaboration \cite{moriya}. The energy dependence
is shown in the panel (d) with data from the LEPS
Collaboration~\cite{niiyama}. The contributions of the contact
terms and the respective meson exchanges are displayed in (b) and
(d) with same nations as in Fig. \ref{fig2}. } \label{fig3}
\end{figure}

The dependences of differential cross sections on the angle and
energy are presented in Figs. \ref{fig3}. The slope of the CLAS
data in the forward direction is reproduced to a degree in the
panels (a), (b), and (c). The rise of the cross section data in
the backward angle in (c) may signify the contributions of the
baryon resonances. For the energy dependence of the differential
cross section in (d) our prediction also agrees with the LEPS data
as well. The contributions of the contact term and the respective
meson exchanges are analyzed in the panels (b) and (d).

\subsection{$\gamma n\to K^+\Sigma^{*-}$}

\begin{figure}[]
\centering \vspace{0.5cm}
\includegraphics[width=0.9\hsize,angle=0]{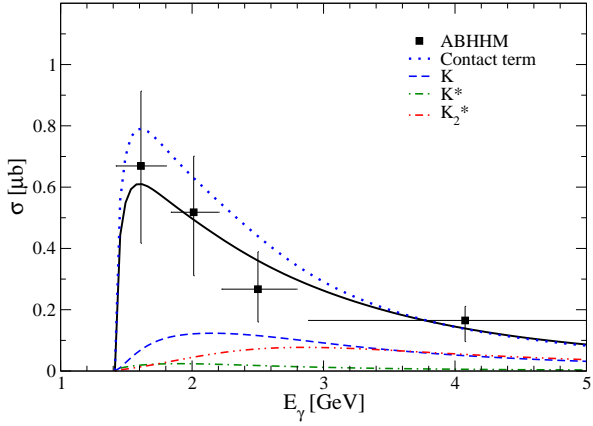}
\caption{Total cross section for $\gamma n \to K^+
\Sigma^{*-}(1385)$. Contributions of the contact term and meson
exchanges are displayed with the same notations as in Fig.
\ref{fig2}. The dominance of the contact term is shown. Data are
taken from Ref. \cite{benz}. } \label{fig4}
\end{figure}

\begin{figure}[]
\centering
\bigskip
\bigskip
\includegraphics[width=0.9\hsize,angle=0]{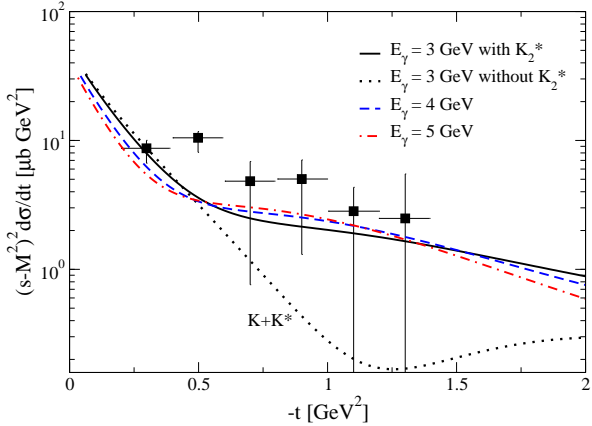}
\caption{Differential cross section $(s-M^2)^2{d\sigma/dt}$ for
$\gamma n \to K^+ \Sigma^{*-}(1385)$ at $E_\gamma=3,\,4$, and 5
GeV. The solid and dotted curves are the cross sections at
$E_\gamma=3$ GeV with and without $K_2^*$, showing the  role of
the tensor meson. Data are taken from Ref.~\cite{benz}. }
\label{fig5}
\end{figure}

There are various sorts of data on the $\gamma n\to
K^+\Sigma^{*-}$ process in comparison to the former $\gamma p$
process. The total and differential cross sections are found in
the experiment at the ABHHM Collaboration in the mid-1970's
\cite{benz}. Very recently the angular distribution and beam
polarization asymmetry were measured in the LEPS experiment
\cite{hicks}.

We calculate  the energy dependence of the cross section  and
present the result in Fig. \ref{fig4}. There might be a room for
improving the accuracy in future experiment as can be seen in Fig.
\ref{fig2}. But the data of the ABHHM are enough  to test our
model prediction at the present stage, exhibiting the maximum peak
and the slope of the decrease along with the increase of photon
energy. We note that the $K_2^*$ exchange gives an equal amount of
contribution to the $K$ over $E_\gamma\approx 3$ GeV.

Figure \ref{fig5} shows the  differential cross section scaled by
the factor $(s-M_n^2)^2$ so that the $-t$ distribution of the
cross section is energy independent. We reproduced the cross
section at the photon energies, 3, 4, and 5 GeV up to the limit of
the experiment, $E_\gamma=5.3$ GeV. It should be pointed out that
the role of the $K_2^*$ is crucial to meet with the data in the
region $-t>0.5$ GeV$^2$/$c^2$.

\begin{figure}[]
\centering
\includegraphics[width=0.9\hsize,angle=0]{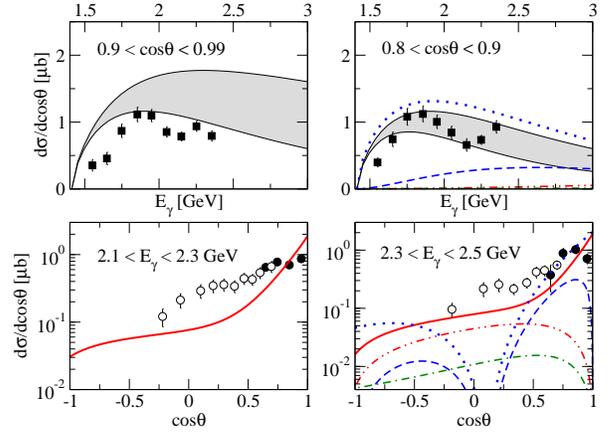}
\vspace{0.5cm} \caption{Dependence of differential cross sections
for $\gamma n \to K^+ \Sigma^{*-}(1385)$ on the energy (upper
panels) and angle (lower panels). Model predictions are given by
the grey band to cover the range of angle denoted. The
contributions of the contact terms and the respective meson
exchanges estimated at $\cos\theta=0.85$ are displayed in upper
right and at $E_\gamma=2.4$ GeV in the lower right panels with
same nations as in Fig. \ref{fig4}. The dip structure appears
there due to the canonical phase of the $K$ exchange. Data of the
LEPS (black squares) are taken from Ref.~\cite{hicks} and those of
the CLAS (empty circles) are from Ref. \cite{mattione}.}
\label{fig6}
\end{figure}

Shown in Fig. \ref{fig6} is the energy dependence of the
differential cross section at forward angles  and its angle
dependence in two energy bins. The energy dependence of the
$d\sigma/d\cos\theta$ is shown in the range calculated between two
boundaries $\cos\theta=0.9$ and $0.99$ in the first panel, for
instance. The angle dependence of $d\sigma/d\cos\theta$ is
calculated at the $E_\gamma=2.2$ and 2.4 GeV, respectively. These
results reproduce quite well the overall feature of the cross
section data. The contributions of the contact terms and the
respective meson exchanges  are analyzed in upper right and lower
right panels, where the dip structure of the  $K$ exchange, and of
the contact term, as a result, are shown at the $-t\approx 0.3$
GeV$^2$ due to the zero of the trajectory  $\alpha_K(t)=0$ in the
canonical phase of the $K$ exchange.

\begin{figure}[]
\centering
\includegraphics[width=0.75\hsize,angle=0]{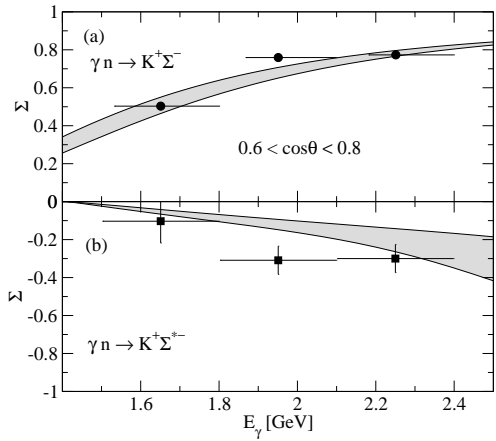}
\caption{Energy-dependence of beam polarization asymmetries for
$\gamma n \to K^+ \Sigma^-(1190)$ from
$\Sigma={d\sigma_y-d\sigma_x\over d\sigma_y+d\sigma_x}\,$(a) and
$\gamma n \to K^+ \Sigma^{*-}(1385)$ from
$\Sigma=-{d\sigma_y-d\sigma_x\over d\sigma_y+d\sigma_x}\,$ (b).
The beam polarization asymmetry in (a) is calculated by the model
in Ref. \cite{bgyu-kaon}. Model predictions are given by the grey
band to cover the range of the angle denoted. Data are taken from
Ref.~\cite{hicks}.} \label{fig7}
\end{figure}

The energy dependence of the beam polarization asymmetry $\Sigma$
was measured in the LEPS experiment of the reaction $\gamma n \to
K^+ \Sigma^{*-}(1385)$ and the result is compared with the case of
the $\gamma n\to K^+\Sigma^-(1190)$ in Fig. \ref{fig7}~ in the
same range of the angle, $0.6<\cos\theta<1$.

With the $\Sigma$ defined as
\begin{equation}\label{beam}
\Sigma = \frac{d\sigma_y - d\sigma_x}{d\sigma_y + d\sigma_x}\,,
\end{equation}
where $d\sigma_{x(y)}$=${d\sigma_{x(y)}\over d\Omega}$ is the
component of the  differential cross section in the $xyz$-system
spanned by the photon momentum ($z$-direction) and two other axes
orthogonal to it in the production plane, we calculated the
$\Sigma$ of the $\gamma p\to K^+\Sigma$ process in Fig. \ref{fig7}
(a) by using the model of Ref. \cite{bgyu-kaon} where the
production amplitude consists of the $K+K^*+K_2^*$ similar to Eq.
(\ref{n+}), but the phases for all the exchanged mesons are taken
to be constant, i.e., 1.
As for the case of the $\gamma n\to K^+\Sigma^{*-}$ process,
however, the beam polarization $\Sigma$ as shown by the grey band
in Fig. \ref{fig7} (b) is predicted in the present framework with
the sign of the $\Sigma$ in Eq. (\ref{beam}) reversed.
At the present stage, we leave it a problem how to reconcile the
sign of the $\Sigma$ between theory and experiment, and suggest
that such an uncertainty in measuring the  $\Sigma$ in the $\gamma
n$ reaction needs to be more analyzed in future experiments.

\section{Summary and Discussion}

In this work, we have investigated the reaction processes $\gamma
p\to K^+\Sigma^{*0}$ and $\gamma n\to K^+\Sigma^{*-}$ to analyze
the production mechanism based on the data provided by the CLAS
and LEPS Collaborations as well as those by the CBCG, ABBHHM, and
ABHHM Collaborations. By using a set of coupling constants common
in both reactions total and differential cross sections as well as
the beam polarization asymmetry are analyzed and the results in
these reactions are quite reasonable to account for the
experimental data. Nevertheless, we need to work further on the
beam polarization $\Sigma$ to resolve the inconsistency between
the model calculation and measurement.

The results obtained in this work show that the most important
contribution comes from the contact term which is a feature of the
spin-3/2 baryon photoproduction. Then, the contribution of the
pseudoscalar $K$ exchange  follows as the dominant one among the
$t$-channel meson exchanges.
The role of the $K^*$ exchange from the present analysis turned
out to be of secondary importance, as concluded in previous works.
Nevertheless, it cannot be neglected in these processes because of
its relation with the $K_2^*$ which plays the role crucial to
explain the data at high energy, as demonstrated in the scaled
differential cross section of the $\gamma n$ reaction.

A few remarks are in order. First, we note that the size of the
total cross section for the $\gamma p$ process is about the same
as that  of the $\gamma n$, though the amplitude of the latter
process differs by a factor of $\sqrt{2}$ from the former, i.e.
\begin{eqnarray}
{\sigma_{\gamma n}\over\sigma_{\gamma p}}\sim {|\sqrt{2}({\rm
contact}+ t{\rm-ch.}\ K+\cdots)|^2\over |{\rm contact}+t{\rm-ch.}\
K+\cdots|^2}\sim1.
\end{eqnarray}
This could be understood as the similar size of the contact term
contribution which is dominant in both reactions, as shown in
Figs. \ref{fig2} and \ref{fig4}.

By comparing the maximum size of the cross section $\sigma\approx
10$ $\mu$b for the $\gamma p\to \pi^+\Delta^{0}$ process \cite{wu}
with that of $\sigma\approx 1$ for the $\gamma p\to
K^+\Sigma^{*0}$ process, their ratio is basically consistent with
the reduction of the leading coupling constant $f_{KN\Sigma^*}$ by
a factor of 36 $\%$ as compared to the $f_{\pi N\Delta}=1.7$ in
the same mass unit, i.e.,
\begin{eqnarray}
{\sigma(\gamma p\to K^+\Sigma^{*0})\over\sigma(\gamma p\to
\pi^+\Delta^{0})}\approx {\left|f_{KN\Sigma^*}\over f_{\pi
N\Delta}\right|^2}.
\end{eqnarray}
Therefore, it is reasonable to assume that both reactions  share
the same production mechanism as the members of the
baryon-decuplet within the present framework.

Finally, we give a comment on the study of $N^*$ resonances,
though it is beyond the scope of the present work. For future work
it is desirable to investigate the role of $N^*$  in the neutral
processes such as in Eqs. (\ref{p0}) and (\ref{n0}), because they
have only $K^*+K_2^*$ exchanges in the $t$-channel which are
expected to be small as can be seen in Figs. \ref{fig2} and
\ref{fig4}. In this sense, the reaction $\gamma p\to
K^0\Sigma^{*+}$ in Eq. (\ref{p0}), in particular, could provide a
ground  more advantageous to identify $N^*$ resonances in the
measured cross section of $\sigma=0.68\pm0.48\, (\mu b)$ at
$E_\gamma=1.42\sim2$ GeV and $\sigma=0.13\pm0.09\, (\mu b)$ at
$E_\gamma=2\sim5.8$ GeV \cite{erbe-pr}, which is of the same order
of magnitude  as the charged ones we have presented in this work.

\acknowledgments

We are grateful to Hungchong Kim for fruitful discussions.  This
work was supported by the grant NRF-2013R1A1A2010504 from National
Research Foundation (NRF) of Korea.

\appendix
\section{SU(3) relation of the meson-baryon coupling constants for
the interactions of the ${\bf 8-8-8}$ and ${\bf 10-8-8}$ types}

We use the phase and coupling constants of the meson-baryon
interaction ($MBB$) of the type ${\bf 8-8-8}$ which is defined by
the following tensor operators,
\begin{eqnarray}
B_i^j=\left(\begin{array}{ccc}
    {1\over \sqrt{2}}\Sigma^0+{1\over \sqrt{6}}\Lambda & \Sigma^+ & p\\
    \Sigma^- & -{1\over \sqrt{2}}\Sigma^0+{1\over \sqrt{6}}\Lambda & n\\
    -\Xi^- & \Xi^0 & -{2\over \sqrt{6}}\Lambda\\
    \end{array} \right)
\end{eqnarray}
for the $J^P={1\over 2}^+$ baryon octet, and
\begin{eqnarray}
M_i^j=\left(\begin{array}{ccc}
    {1\over \sqrt{2}}\pi^0+{1\over \sqrt{6}}\eta & \pi^+ & K^+\\
    \pi^- & -{1\over \sqrt{2}}\pi^0+{1\over \sqrt{6}}\eta & K^0\\
    K^- & \bar{K}^0 & -{2\over \sqrt{6}}\eta\\
    \end{array} \right)
\end{eqnarray}
for the $J^P=0^-$ pseudoscalar meson octet.

The meson-baryon-baryon ($MBB$) interaction of the ${\bf 8-8-8}$
type can be constructed from fully contracting the indices as
\begin{eqnarray}\label{app1}
a\bar{B}^i_j B^j_k M^k_i  + b\bar{B}^i_j B^k_i M^j_k + {\rm h.c.},
\end{eqnarray}
Therefore, two types of coupling are possible in the SU(3) limit,
which are equivalent to the conventional $F$ and $D$ types.

For the $J^P={3\over 2}^+$ baryon decuplet, totally symmetric
tensor $D^{ijk}$ can be identified with the baryon resonances.
\begin{widetext}
\begin{eqnarray}
D^{111}=\Delta^{++} ,\,\,\, D^{112}={1\over\sqrt{3}}\Delta^+,
\,\,\,
D^{122}={1\over\sqrt{3}}\Delta^0, \,\,\, D^{222}=\Delta^- ,\\
D^{113}={1\over\sqrt{3}}\Sigma^{*+}, \,\,\,
D^{123}={1\over\sqrt{6}}\Sigma^{*0}, \,\,\,
D^{223}={1\over\sqrt{3}}\Sigma^{*-} \label{223},\\
D^{133}={1\over\sqrt{3}}\Xi^{*0}, \,\,\,
D^{233}={1\over\sqrt{3}}\Xi^{*-}, \\
D^{333}=\Omega^{*-}.
\end{eqnarray}
\end{widetext}

The meson-baryon-decuplet baryon ($MBD$) interaction of the ${\bf
10-8-8}$ type in SU(3) limit can be again from fully contracting
the indices as
\begin{eqnarray}\label{app2}
g\bar{D}^{ijk}B^l_j M^m_k \epsilon_{ilm} + {\rm h.c.},
\end{eqnarray}
where the Levi-Civita tensor $\epsilon_{ilm}$ is needed because
the total number of index is odd. Therefore, only one type of
coupling is possible in the SU(3) limit as in Eq. (\ref{app2}).

After a little algebra, the following relation is obtained;
\begin{eqnarray}\label{a8}
\frac{f_{\pi^- p \Delta^{++}}}{m_\pi}&=&-\sqrt{6}
\frac{f_{K^+p\Sigma^{*0}}}{m_K}=-\sqrt{3}
\frac{f_{K^+n\Sigma^{*-}}}{m_K}\nonumber\\
&=& \sqrt{3} \frac{f_{K^0p\Sigma^{*+}}}{m_K}=\sqrt{6}
\frac{f_{K^0n\Sigma^{*0}}}{m_K}\,.
\end{eqnarray}


\end{document}